\documentclass[aps,prl,twocolumn,superscriptaddress]{revtex4-1}

\usepackage{graphicx}

\begin{document}

\title{Manipulating domain wall chirality by current pulses in Permalloy/Ir nanostrips}
\author{Muhammad Zahid Ishaque}
\affiliation{CNRS, Institut N\'{e}el, 38042 Grenoble, France} \affiliation{Univ.~Grenoble Alpes, Institut N\'{e}el, 38042 Grenoble, France} \affiliation{University of Sargodha, 40100, Sargodha, Pakistan}
\author{Van Dai Nguyen}
\affiliation{CNRS, Institut N\'{e}el, 38042 Grenoble, France} \affiliation{Univ.~Grenoble Alpes, Institut N\'{e}el, 38042 Grenoble, France}
\author{Olivier Fruchart}
\affiliation{CNRS, Institut N\'{e}el, 38042 Grenoble, France} \affiliation{Univ.~Grenoble Alpes, Institut N\'{e}el, 38042 Grenoble, France}
\author{Stefania Pizzini}
\affiliation{CNRS, Institut N\'{e}el, 38042 Grenoble, France} \affiliation{Univ.~Grenoble Alpes, Institut N\'{e}el, 38042 Grenoble, France}
\author{Nicolas Rougemaille}
\affiliation{CNRS, Institut N\'{e}el, 38042 Grenoble, France} \affiliation{Univ.~Grenoble Alpes, Institut N\'{e}el, 38042 Grenoble, France}
\author{Svenja Perl}
\affiliation{CNRS, Institut N\'{e}el, 38042 Grenoble, France} \affiliation{Univ.~Grenoble Alpes, Institut N\'{e}el, 38042 Grenoble, France}
\author{Jean-Christophe Toussaint}
\affiliation{CNRS, Institut N\'{e}el, 38042 Grenoble, France} \affiliation{Univ.~Grenoble Alpes, Institut N\'{e}el, 38042 Grenoble, France}
\author{Jan Vogel} \email{jan.vogel@neel.cnrs.fr}
\affiliation{CNRS, Institut N\'{e}el, 38042 Grenoble, France} \affiliation{Univ.~Grenoble Alpes, Institut N\'{e}el, 38042 Grenoble, France}

\begin{abstract} Using magnetic force microscopy and micromagnetic simulations, we studied the effect of Oersted magnetic fields on the chirality of transverse magnetic domain walls in Fe$_{20}$Ni$_{80}$/Ir bilayer nanostrips. Applying nanosecond current pulses with a current density of around $2\times10^{12}$ A/m$^2$, the chirality of a transverse domain wall could be switched reversibly and reproducibly. These current densities are similar to the ones used for current-induced domain wall motion, indicating that the Oersted field may stabilize the transverse wall chirality during current pulses and prevent domain wall transformations.\end{abstract}

\maketitle

When a longitudinal electrical current is passing through a magnetic nanostrip containing a magnetic domain wall (DW), both the spin and the charge of the conduction electrons act on the DW magnetization. In single ferromagnetic layers, the spin polarization of the current induces a local torque on the domain wall, called the spin-transfer torque (STT) \cite{Berger1984,Slonczewski1996}. In multilayered structures containing heavy metals like Pt or Ta supplementary torques induced by the spin-orbit coupling also influence the magnetization \cite{Miron2010,Liu2012}. On the other hand, the charge of the current induces an Oersted magnetic field transverse to the current direction. This Oersted field increases with the distance from the center of the current flow and is proportional to the current density. For strips with a single metallic layer, the in-plane transverse component of the Oersted field can usually be neglected for thin enough strips since the net Oersted field is zero and the maximum fields acting in opposite directions at the bottom and top interfaces of the strip are small. As the Oersted field has a component perpendicular to the plane of the strip close to the borders, in materials with perpendicular magnetization it can induce a tilted domain wall shape \cite{Adam2009} or even a bi-domain state with a DW parallel to the strip axis \cite{Boulle2009}. In asymmetric strips with in-plane magnetization having different buffer and capping layers, but especially in ferromagnetic(FM)/non-magnetic(NM) bilayer or FM/NM/FM trilayer strips, a significant net transverse Oersted field can exist and influence the magnetization configuration of the FM layer(s) during current pulses \cite{Uhlir2011}. In particular, it may influence the configuration of domain walls that are present in the strip, favoring transverse domain walls (TW) with a magnetization parallel to the field. The stabilization of one particular TW configuration by a transverse (Oersted) field during field or current-induced DW motion (CIDM) may delay the onset of domain wall transformations associated to the Walker breakdown \cite{Walker1974} up to higher domain wall velocities \cite{Bryan2008,Glathe2008b,Kunz2008,Jang2010,Jang2012,Goussev2013}.

In this work, we have used Magnetic Force Microscopy (MFM) and micromagnetic simulations to study the influence of current pulses on the configuration of magnetic domain walls in Fe$_{20}$Ni$_{80}$(Py)/Ir bilayer nanostrips with similar Py and Ir thickness. We show that the chirality of the transverse DW can indeed be manipulated with current pulses, and we determine the current density needed for switching the chirality of transverse domain walls as a function of strip width and Ir thickness. The associated switching fields are compared to the results of micromagnetic simulations. The current densities for which chirality switching takes place are similar to the ones needed for current-induced domain wall motion, indicating that the Oersted field should have an important influence on CIDM in this type of multilayer systems.

\begin{figure}[ht!]
\includegraphics[width=8cm]{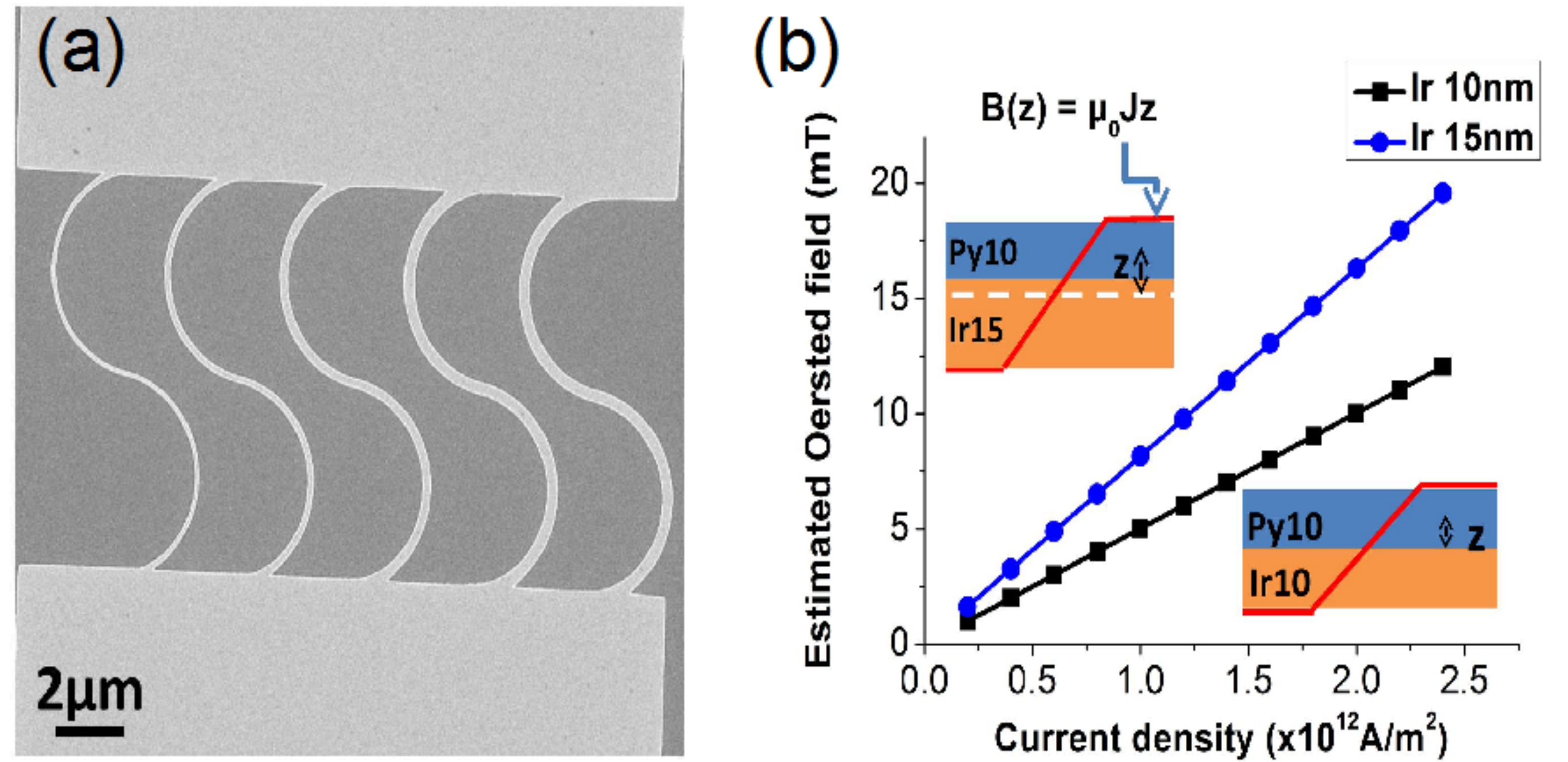}
\caption{\label{SEMOersted} (a) SEM image (Topography) of 20 $\mu$m long, 100 to 400~nm wide S-shaped nanostrips together with the injection pads (b) In-plane Oersted field (B) as a function of current density for 10 nm of Py and Ir thicknesses of 10 and 15 nm. The field value corresponds to the average field acting on the Py layer, i.e. the field in the center of the Py layer. The red lines in the insets show the Oersted field profile across the Py thickness and the dotted white line indicates the center of the Py(10nm)/Ir(15m) structure.}
\end{figure}

Samples with two different Ir thicknesses were studied, Py(10 nm)/Ir(10 nm) and Py(10 nm)/Ir(15 nm), epitaxially grown on sapphire(0001) \cite{ZahidUnpub} and capped by 2~nm of Au. The different thicknesses of Ir allow obtaining different values of the Oersted field acting on the Py layer for a same current density. Nanostrips with lengths of 10 and 20 $\mu$m and 100 to 400~nm wide were patterned using electron-beam lithography and ion-beam etching. Fig.~\ref{SEMOersted}(a) shows the scanning electron microscope (SEM) image of the S-shaped nanostrips, which are connected to Au contacts through the two injection pads. TWs were created at the bends of these nanostrips by applying a 50~mT magnetic field transverse to the strips. Subsequently, current pulses of densities between 1.5 and 3$\times$10$^{12}$~A/m$^2$ and pulse lengths of 3~ns, with rise and fall times of 0.8 ns, were applied. This pulse length was chosen to be longer than typical precession times and short enough to limit the influence of thermal effects on the DW configuration. For estimating the amplitude of the Oersted field induced by the current pulses, we supposed in a first approximation that the resistivities of Au, Py and Ir and therefore the current densities in the layers were the same. As the Oersted field is not uniform over the Py thickness, we considered the average Oersted field, i.e. the value calculated at the center of the Py layer. This value was obtained using the relation B(z) = $\mu_0$Jz, where J is the current density and z is the vertical distance from the center of the Au/Py/Ir trilayer structure [Fig.~\ref{SEMOersted}(b)].

\begin{figure}[ht!]
\centerline{\includegraphics[width=6cm]{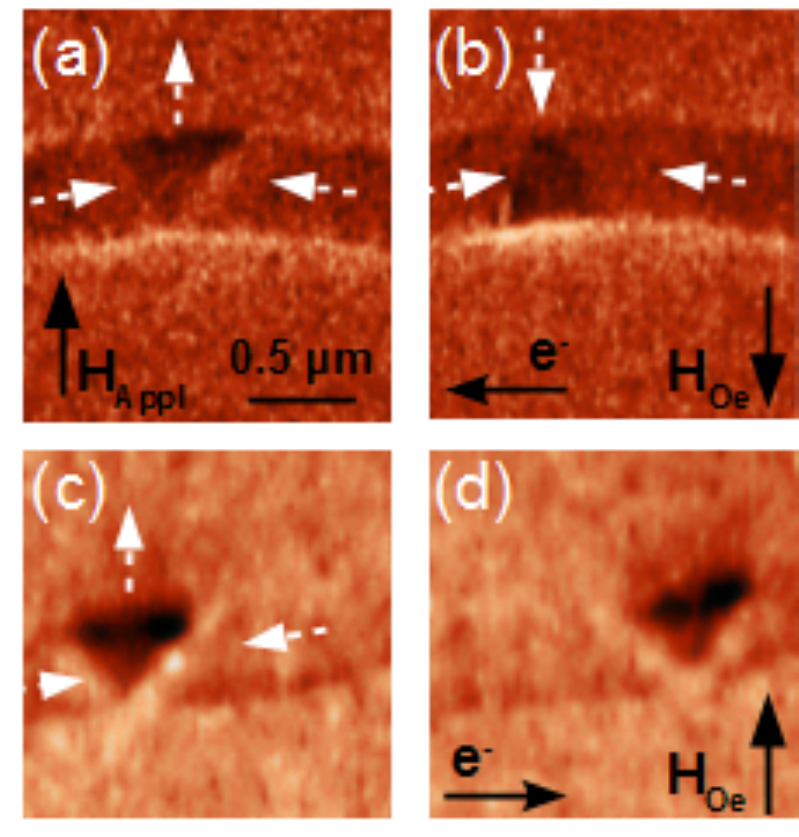}}
\caption{\label{TWs} (a,b) Transverse domain wall manipulation by current pulses in a 20 $\mu$m long and 400~nm wide S-shaped Au(2nm)/Py(10nm)/Ir(15nm)/Al$_2$O$_3$(0001) nanostrip. MFM images of (a) the initial configuration of a head-to-head TW and (b) after applying a 1.6~$\times$~10$^{12}$ A/m$^2$, 3~ns current pulse. (c) and (d) Displacement of a TW in a 400~nm wide Au(2nm)/Py(10nm)/Ir(10nm)/Al$_2$O$_3$(0001) nanostrip induced by a 3~ns current pulse with a current density of 2.4~$\times$~10$^{12}$ A/m$^2$. Both the initial and final magnetization direction of the TW are parallel to the Oersted field direction. The dotted white arrows indicate the magnetization direction in the strip and the domain walls, while the black arrows give the direction of electron flow and the directions of applied and Oersted fields.}
\end{figure}

MFM images were taken in air and at 300K, using a NT-MDT microscope NTegra-Aura with custom-made low moment magnetic tips, obtained by depositing 3-5 nm of Co$_{80}$Cr$_{20}$ on commercial AC240TS probes from Olympus. Fig.~\ref{TWs}(a,b) shows an example of the TW switching, for a 400~nm wide Py(10nm)/Ir(15nm) nanostrip. In the initial state the TW has its magnetization pointing up, as imposed by the applied magnetic field (Fig.~\ref{TWs}(a)). After a pulse of duration 3~ns and current density 1.6$\times$10$^{12}$ A/m$^2$, corresponding to an Oe field of about 10 mT pointing down, the TW magnetization has switched (Fig.~\ref{TWs}(b)).

\begin{figure}[ht!]
\centerline{\includegraphics[width=7cm]{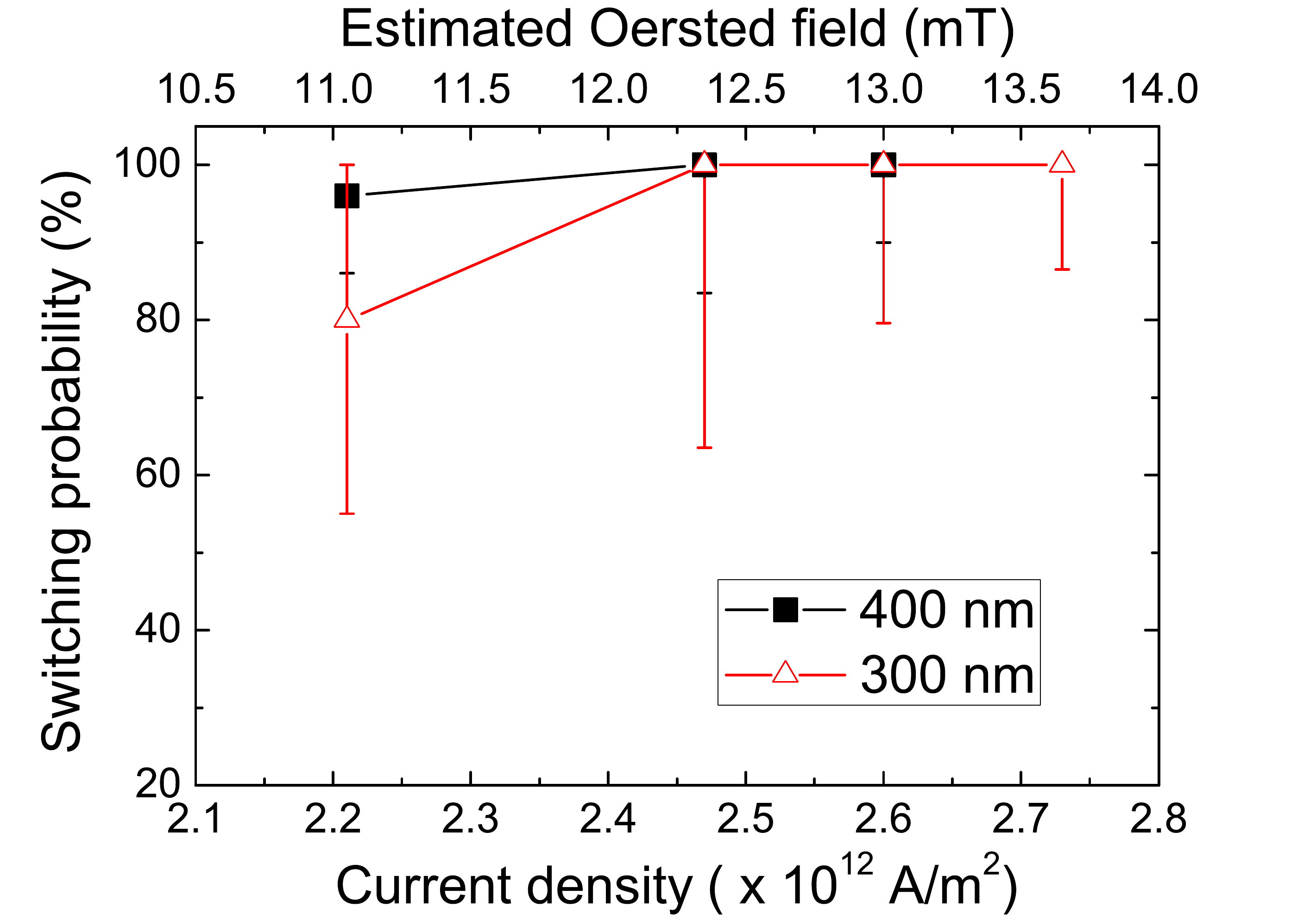}}
\caption{\label{switching} TW switching probability as a function of current density and corresponding Oersted field in 300 and 400~nm wide Au(2nm)/Py(10nm)/Ir(10nm)/Al$_2$O$_3$(0001) nanostrips.}
\end{figure}

The probability of DW chirality switching was measured for different current densities. Fig.~\ref{switching} shows the DW switching probability as a function of current density and corresponding Oersted field for Py(10 nm)/Ir(10 nm) bilayers. The DW chirality switching probability increases with increasing current density. Current densities between 2.3 and 2.6$\times$10$^{12}$A/m$^2$ were required to switch the DW chirality with 100$\%$ probability, corresponding to estimated Oersted fields of 11.7$\pm$3~mT and 12.8$\pm$3~mT for the 400 and 300 nm wide strips, respectively. The 300~nm wide TWs show a higher deterministic switching field, as expected due to the higher transverse demagnetizing field for narrower nanostrips. The error bars on the determined experimental switching fields are rather large, since the switching fields are extrapolated from a limited number of points (a maximum of ten events per current density value) and the approximation of a homogeneous current density over the different layers. When increasing the thickness of the Ir layer from 10~nm to 15~nm, lower current densities are needed to switch the DW chirality with 100$\%$ probability (between 1.5 and 1.7$\times$10$^{12}$A/m$^2$), but the estimated Oersted fields for switching are similar : 10.6$\pm$3~mT for 400 nm and 12.5$\pm$3~mT for 300 nm wide strips. We also tried to switch the TW chirality in 100 and 200 nm wide strips. However, no switching was observed with the highest available current density (about 3 $\times$ 10$^{12}$ A/m$^2$).

According to the DW phase diagram \cite{Nakatani2005,Rougemaille2012}, the energetically most favorable DW structure for a Py thickness of 10 nm and strip widths above 150~nm is the vortex wall. Such vortex walls were sometimes observed in the 300 and 400 nm wide strips, after applying pulses with a current density corresponding to a switching probability lower than 100$\%$ in Fig.~\ref{switching}). For higher current densities we always observed TWs, indicating that the energy barrier for the transformation of this metastable state to the more stable VW can not be overcome at room temperature.

In most cases, the TWs switched chirality without changing position, but sometimes a DW displacement was observed upon switching the chirality. Such a displacement can be induced by STT, by inertial auto-motion \cite{Chauleau2010} or by a combination of both. In fact, a displacement of TWs of several micrometers upon transformation to vortex walls has been observed in literature \cite{Chauleau2010}. The fact that we do not observe this kind of auto-motion systematically indicates the important pinning of the DWs in our sample.

It has been reported that, for very short current pulses, the TW chirality can also switch under the effect of STT \cite{Vanhaverbeke2008,Tretiakov2012,Sedlmayr2013}. In order to exclude that the TW chirality can switch also against the direction of the Oersted field, we applied current pulses with Oersted fields parallel to the TW magnetization. In that case, DW transformations were not observed. Moreover, despite the strong DW pinning, we sometimes observed TW motion without transformations, along the electron flow, when the Oersted field was parallel to the TW magnetization (Fig.~\ref{TWs}(c,d)). This behavior confirms that the Oe field tunes and stabilizes the chirality of TWs.

\begin{figure}[ht!]
\centerline{\includegraphics*[width=7cm]{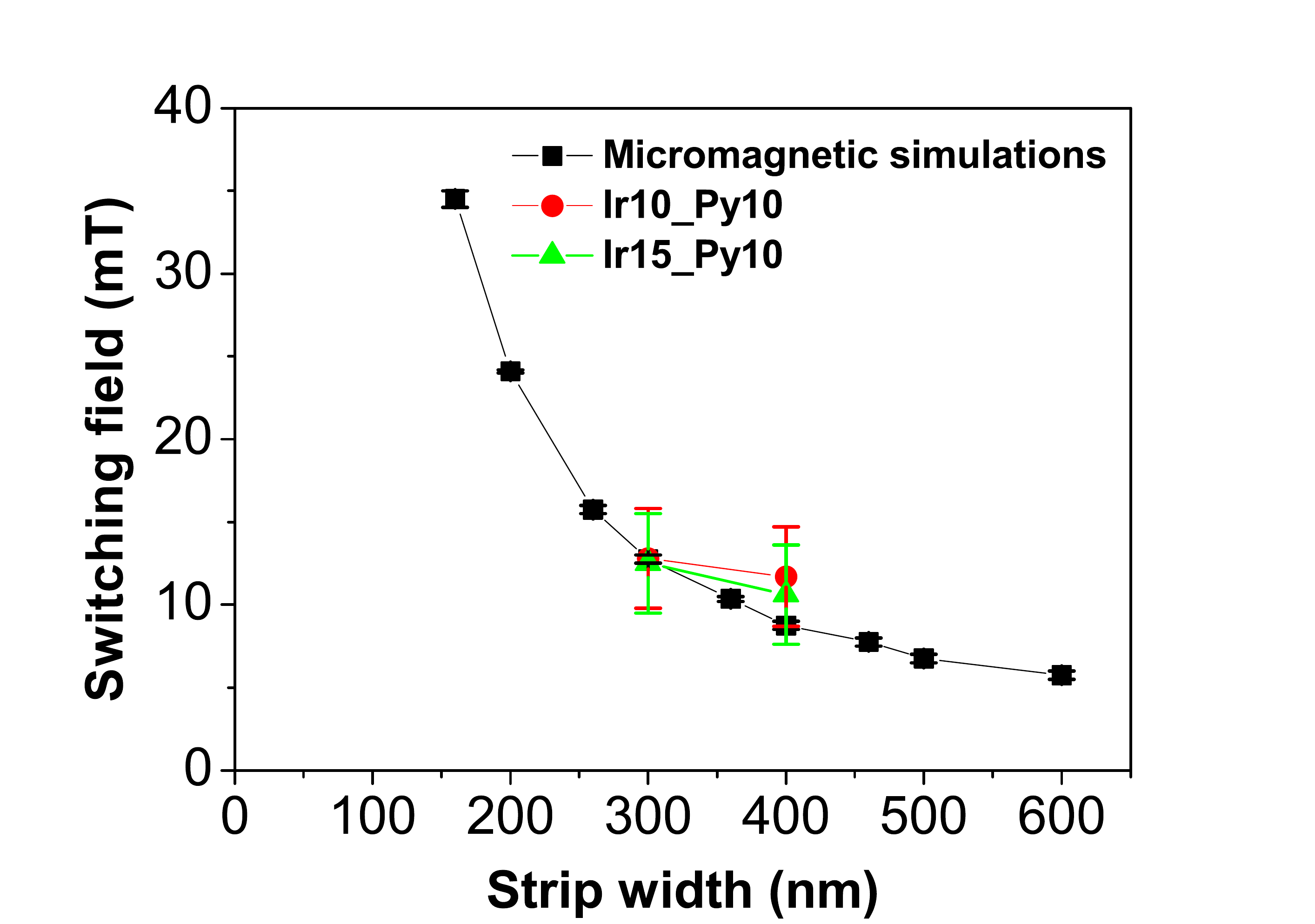}}
\caption{\label{simTW} Simulated switching field for a TW in 10~nm wide Py nanostrips with different widths (black squares). A comparison with the experimental switching fields for 10 (red dots) and 15 nm (blue triangles) of Ir is also given.}
\end{figure}

In order to compare the experimental values found for the TW switching fields with a model case, we performed micromagnetic simulations using the finite differences OOMMF code from NIST \cite{OOMMF}. In these simulations, spontaneous magnetization and exchange stiffness were set to $\mu_{0}M_{Py}$~=~1.0053 T and $A_{Py}$~=~10 pJ/m, respectively. Magnetocrystalline anisotropy was zero, while the cell size was set to $4\times 4\times 1$ nm$^3$ and the damping parameter to 0.01. The thickness of the Py layer was fixed to 10 nm and its width was varied from 160 to 600 nm. To minimize size effects, the length was chosen to be at least 20 times the width, and magnetic moments at the extremities of the nanostrips were fixed to avoid non-uniform magnetization profiles at the edges. Initial conditions were such that a relaxed TW is located in the center of the nanostrips. A homogeneous magnetic field step was then applied from remanence, with only a component transverse to the nanostrip, and the system was let to evolve. If switching did not occur, the amplitude of the magnetic field step was increased by 0.25~mT each time (starting again from remanence), until chirality switching was observed. Results from these simulations are reported in Fig.~\ref{simTW} which compares the simulated switching fields with the experimental results. The switching field is found to scale with the inverse of the nanostrip width, and a good general agreement between the simulations and the experimental results is found.

Our measurements show that in Py(10m)/Ir(10nm,15nm) bilayer nanostrips the current density needed to impose a TW chirality is of the order of 1.5 to 2.5~$\times$~10$^{12}$~A/m$^2$, corresponding to Oersted fields of 10 to 15~mT. These current density values are of the same order of magnitude as those reported in the literature for domain wall propagation in single layer Py nanostrips. We therefore expect  the Oe field to have an influence on DW motion through the stabilisation of the DW chirality. In fact, it has been shown by Eastwood \textit{et al.} \cite{Eastwood2011} that for Py nanostrips with similar width (390~nm) and thickness (10~nm) the TW chirality can already be stabilized with a transverse field of the order of 5~mT. In previous measurements on trilayer Py/Cu/Co nanostrips, we observed high current-induced DW velocities for current densities corresponding to transverse Oersted fields of about 3-4~mT \cite{Uhlir2010}. In that case, the Py layer thickness was smaller (5nm), leading to a higher stability of the TW configuration according to the DW phase diagram \cite{Nakatani2005}, a stability that is further increased by the magnetostatic interaction with the Co layer \cite{Rougemaille2012}. The stabilization of a given TW configuration and the consequent shift of the Walker breakdown to higher current densities may contribute to the high maximum domain wall observed in these trilayer nanostrips \cite{Uhlir2010,Pizzini2009}.

Fixing the chirality of a transverse domain wall by an Oersted field can also be of interest in other type of bilayer strips. It was shown in a recent theoretical work that in Py/Pt bilayer strips TWs can move through a combination of STT and vertical spin currents coming from the Pt layers, due to the spin Hall effect \cite{Seo2012}. The direction of motion of the TW depends on its chirality with respect to the polarity of the spin Hall current. Very high domain wall velocities against the electron flow direction are expected, for current densities just below the threshold current for spin Hall current induced TW switching. If the current density for TW switching can be increased by adding an Oersted field, higher maximum domain wall velocities may be reached. Engineering hybrid systems with in-plane magnetization that lead to a combination of current-induced torques acting on the domain wall magnetization, like STT, spin Hall current and Oersted fields, may lead to much higher current-induced domain wall velocities, as it was already shown for perpendicular magnetic systems \cite{Miron2011,Ryu2013,Emori2013}.

This work was partially supported by the Agence Nationale de la Recherche through project ANR-11-BS10-008 Esperado and by the Fondation Nanosciences. Z.M.I thanks the Higher Education Commission (HEC) of the government of Pakistan for a Ph.D. grant. Sample patterning was performed at the Nanofab facility of the Institut N\'{e}el. We thank S.~le~Denmat, P.~David and V.~Guisset for experimental help.


\begin{thebibliography}{9}
\bibitem{ZahidUnpub} Details on the epitaxial growth and sample preparation will be published in a forthcoming paper.
\end{thebibliography}
\end{document}